# XOR AND XNOR GATES IN INSTANTANEOUS NOISE BASED LOGIC

MOHAMMAD B. KHREISHAH [1], WALTER C. DAUGHERITY [2], LASZLO B. KISH [1,3]

[1] *Department of Electrical and Computer Engineering, Texas A&M University, College Station, TX 77843-3128, USA*

[2] *Department of Computer Science & Engineering, Texas A&M University, College Station, TX 77843-3112, USA*

[3] *Obuda University, Budapest, Bécsi út 96/B, Budapest, H-1034, Hungary*

In this paper, we propose a new method of applying the XOR and XNOR gates on exponentially large superpositions in Instantaneous Noise-Based Logic. These new gates are repeatable, and they can achieve an exponential speed up in computation with a polynomial requirement in hardware complexity.

## 1. Introduction

Noise-based logic (NBL) was first introduced in [1], where orthogonal stochastic processes (noises), their superposition, their products and the superposition of their products are used to represent the logic state [2]. Instantaneous Noise-based Logic (INBL) [3] is a class of NBL, where the signals carrying the logic values appear immediately at the output of the logic gates without the need of cross-correlating or time averaging.

Previously, NOT [3,4] and CNOT [12] gates in INBL were created. This paper proposes the solution for XOR and XNOR gates in INBL.

In the next subsections we show a few details of NBL that are essential for the present paper.

### 1.1. *On Noise-Based Logic*

NBL requires a Reference Noise System (RNS) as the source for generating the logic states and for the identification of the incoming noises that carry the logic values.

A system with *N* noise-bits uses 2*N* independent, orthogonal noise sources for the RNS [1-11]. Let each noise of the RNS be noted by $W_{i,j}(t)$, where [*i*] is the bit significance number ($1 \leq i \leq N$), and [*j*] is the value of that bit number ($j \in \{0,1\}$). Any binary number, *R*, in the range of $[0, 2^N]$, is represented by the product of *N* noise sources $X_R(t)$ which we will call a "string":

$$X_R(t) = \prod_{i=1}^{N} W_{i,j(i,R)}(t) \tag{1}$$

For example, a system of 2 noise-bits will have the following 2*N* independent noise sources:

$$W_{i,j}(t) = \{W_{1,0}(t), W_{1,1}(t), W_{2,0}(t), W_{2,1}(t)\} \tag{2}$$

The following string represents the number 2 that has the binary number representation $(10)_2$:

$$X_{10}(t) = W_{2,1}(t)W_{1,0}(t) \tag{3}$$

In NBL, a superposition of strings simply means the summation of these strings but not the summation of the numbers they represent. Suppose that $Y(t)$ is a superposition of strings, then we have:

$$Y(t) = X_{R1}(t) + X_{R2}(t) + X_{R3}(t) + \cdots \tag{4}$$

Notice a few characteristics of a superposition:

(i) The maximum number of strings in a superposition is $2^N$ and the minimum number of strings is one.

(ii) The number of all possible subspaces of a superposition, $S_{total}$, is:

$$S_{total} = \sum_{k=1}^{2^N} \binom{2^N}{k} = 2^{2^N} - 1 \tag{5}$$

## 1.2. *Instantaneous Noise-Based Logic & Random telegraph waves:*

INBL is a class of the NBL family where cross correlation/time averaging is not required. It has similar logic structure as the quantum computer idea.

The Random Telegraph waves (RTW) [3-12] are the simplest form of the RNS of INBLs.

RTWs are synchronous signals that only change when a new clock cycle begins. At the start of the clock cycle, RTWs start randomly either at +1 or at -1, then at the beginning of each new clock period, RTWs have a probability of 0.5 to flip from +1 to -1 or vice versa. RTW has many practical realizations, but we will restrict our discussion to the above form which is the simplest. That means that statistically, RTWs are +1 half the time, and -1 in the other half, which also means that their mean is zero [12]:

$$\langle R_{i,j}(t) \rangle = 0 , \tag{6}$$

where $R_{i,j}(t)$ is RTW with binary significance $i$ and binary value $j$.

The product of RTWs is a new RTW which is orthogonal to each RTW in the product [12]:

$$R_{k,l}(t) = R_{i,j}(t)R_{n,m}(t) \tag{7}$$

$$\langle R_{i,j}(t)R_{k,l}(t)\rangle = \langle R_{n,m}(t)\rangle = 0 \tag{8}$$

$$\langle R_{n,m}(t)R_{k,l}(t)\rangle = \langle R_{i,j}(t)\rangle = 0 \tag{9}$$

Where $i \neq n$ and $j \neq m$.

Also, it is important to note that any RTW multiplied by itself will result in 1 [12]:

$$R_{i,j}(t)R_{i,j}(t) = 1 \tag{10}$$

Equation (10) will be important in applying the XOR/XNOR operation.

### 1.3. *NOT operation in INBL*

Previously, the CNOT and the NOT gates were proposed by acting on the reference wires [12]. Here the NOT gate is reviewed since the new XOR/XNOR gates proposed in the present paper is based on a similar method. From now on, we will be omitting the time notation for convenience, but every signal that is mentioned in this paper is a function of time. Let us suppose that we have the following arbitrary superposition that contains $R_{i0}$ and $R_{i1}$:

$$Y_{total} = Y_0 R_{i0} + Y_1 R_{i1} \quad , \tag{11}$$

where $Y_0$ and $Y_1$ are also superpositions with the restriction that they do not contain $R_{i0}$ or $R_{i1}$. Figure 1 illustrates such a system.

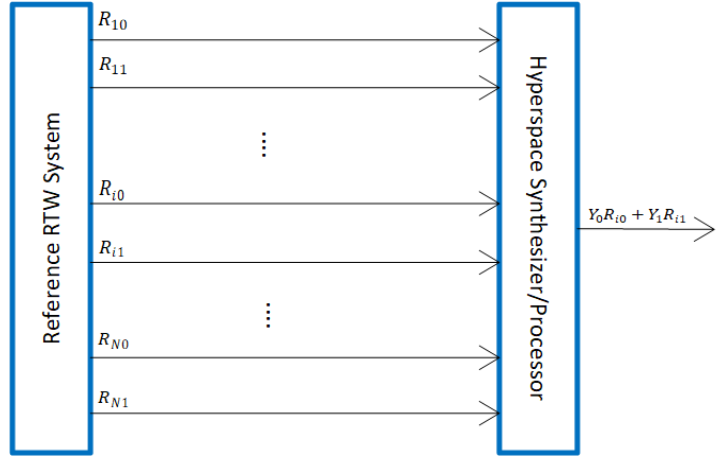

Figure 1: A generic INBL that uses RTW signals and contains $R_{i0}$ & $R_{i1}$ at the output.

To apply the NOT gate on $R_{i0}$ and $R_{i1}$, we simply multiply the reference wires $R_{i0}$ and $R_{i1}$ by $R_{i0}R_{i1}$ [12]. Then due to equation (10), equation (11) becomes:

$$Y_{total}(t) = Y_1 R_{i0} + Y_0 R_{i1} . \qquad (12)$$

Figure 2 illustrates such a system.

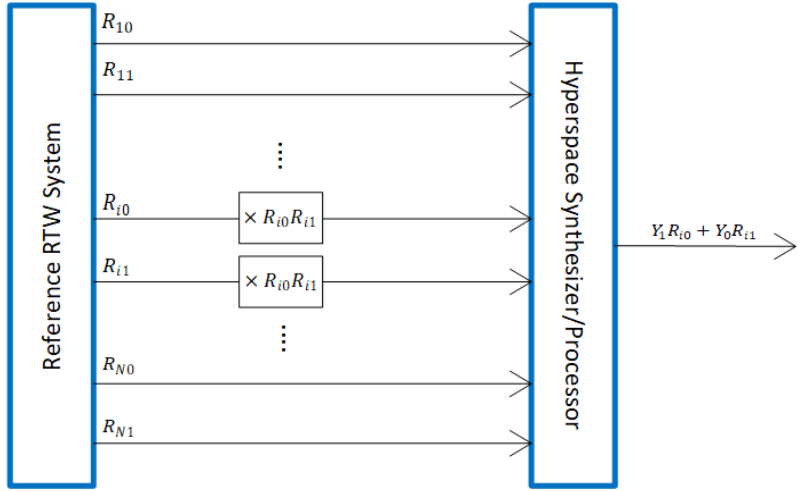

Figure 2: NOT operation in INBL (RTW) system.

## 2. The XOR & XNOR gates in INBL

### 2.1. *The XOR Gate*

Similarly to the NOT gate, we will be using operations on the reference wires to apply the XOR gate. Let us assume that the inputs to the XOR gate are bits $\{i, f\}$, and that we want the result to appear on bit $\{h\}$. Let us assume that we have a processor that produces the following superposition:

$$Y_{total} = Y_0 R_{i0} R_{f0} R_{hx_0} + Y_1 R_{i1} R_{f0} R_{hx_1} + Y_2 R_{i0} R_{f1} R_{hx_2} + Y_3 R_{i1} R_{f1} R_{hx_3} \tag{13}$$

Where $R_{i0}$ represents the zero value of the $ith$ bit; $R_{i1}$ represents the one value of the $ith$ bit; $R_{j0}$ represents the zero value of the $jth$ bit; $R_{j1}$ represents the one value of the $jth$ bit; $R_{hx_0}, R_{hx_1}, R_{hx_2}, R_{hx_3}$ represent the $hth$ bit that has arbitrary initial binary values of $\{x_0, x_1, x_2, x_3\}$, , where these initial values are unimportant;
$Y_0, Y_1, Y_2, Y_3$ are arbitrary superpositions that do not contain any RTW with index $\{i, f, h\}$.
This system is illustrated in Figure 3.

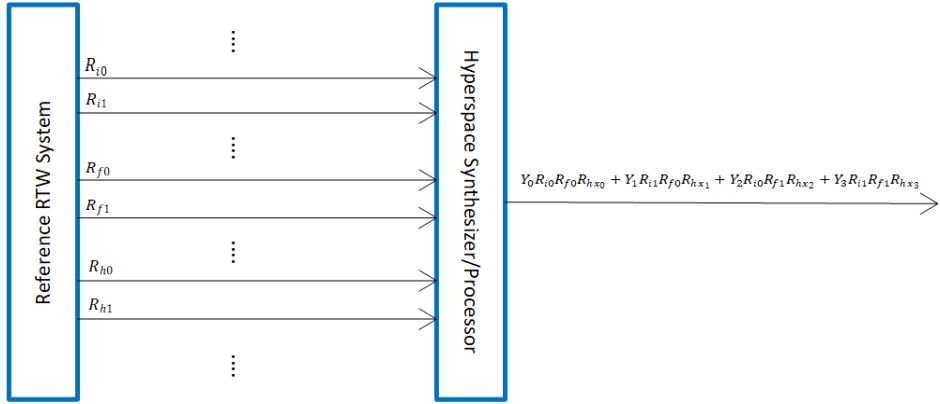

Figure 3: Generic RTW system before applying the XOR/XNOR gates

The first step is to manipulate the superposition shown by equation (13) so that the bit values corresponding to $\{x_0, x_1, x_2, x_3\}$ become 0. This can be done by multiplying the reference wire of $R_{h1}$ by $R_{h0}R_{h1}$. If a string contains $R_{h0}$, then it remains the same, and if it contains $R_{h1}$, then it flips it to $R_{h0}$. This is illustrated by Figure 4. The superposition in equation (13) will then become:

$$Y_{total} = Y_0 R_{i0} R_{f0} R_{h0} + Y_1 R_{i1} R_{f0} R_{h0} + Y_2 R_{i0} R_{f1} R_{h0} + Y_3 R_{i1} R_{f1} R_{h0} \qquad (14)$$

That makes bit $h$ independent of bits $\{i, f\}$. This step is essential because we need the output bit $h$ to start with the same value in each element of the superposition, then to change according to the XOR function of the $\{i, f\}$ bit values, see below.

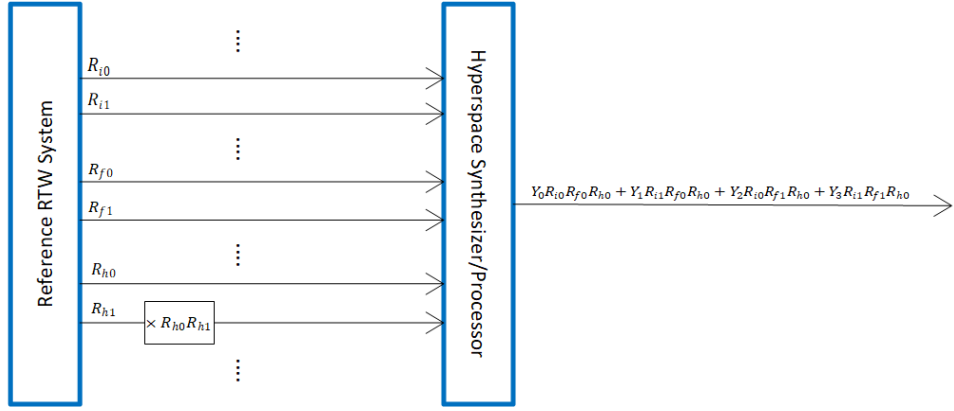

Figure 4: Making the value of bit {h} equals binary value 0 in all strings in the superposition.

The next step is to multiply the $R_{i1}$ and $R_{f1}$ references by $R_{h0}R_{h1}$, which is the main XOR operation. In equation (14), if we substitute $R_{i1}$ by $R_{i1}R_{h0}R_{h1}$ and $R_{f1}$ by $R_{f1}R_{h0}R_{h1}$, then we get the following:

$$Y_{XOR} = Y_0 R_{i0} R_{f0} R_{h0} + Y_1 R_{i1} R_{h0} R_{h1} R_{f0} R_{h0} + Y_2 R_{f1} R_{h0} R_{h1} R_{i0} R_{h0} + Y_3 R_{i1} R_{h0} R_{h1} R_{f1} R_{h0} R_{h1} R_{h0}$$

$$= Y_0 R_{i0} R_{f0} R_{h0} + Y_1 R_{i1} R_{f0} R_{h0} R_{h1} R_{h0} + Y_2 R_{i0} R_{f1} R_{h0} R_{h1} R_{h0} + Y_3 R_{i1} R_{f1} R_{h0} R_{h1} R_{h0} R_{h1} R_{h0}$$

$$= Y_0 R_{i0} R_{f0} R_{h0} + Y_1 R_{i1} R_{f0} NOT(R_{h0}) + Y_2 R_{i0} R_{f1} NOT(R_{h0}) + Y_3 R_{i1} R_{f1} NOT(NOT(R_{h0})) \tag{15}$$

From equations (10) and (15) it follows:

$$Y_{XOR} = Y_0 R_{i0} R_{f0} R_{h0} + Y_1 R_{i1} R_{f0} R_{h1} + Y_2 R_{i0} R_{f1} R_{h1} + Y_3 R_{i1} R_{f1} R_{h0} \tag{16}$$

In conclusion, we successfully implemented the XOR gate between bits $\{i, f\}$ and represented the answer by bit $\{h\}$. Figure 5 illustrates the total XOR gate.

Note, to get exactly the same result, $R_{i0}$ and $R_{f0}$ could have been multiplied by $R_{h0}R_{h1}$ instead of multiplying $R_{i1}$ and $R_{f1}$.

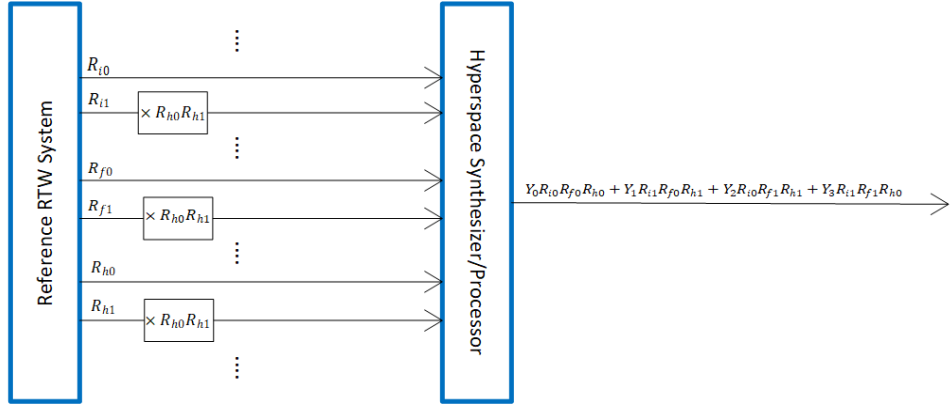

Figure 5: The full XOR gate implementation

## 2.2. The XNOR Gate

Since we implemented the XOR gate, we can simply use the result from XOR gate and the previously developed NOT gate [12] to get the XNOR gate. If we apply the NOT gate to bit $\{h\}$ in equation (16), then we get the XNOR gate:

$$Y_{XNOR} = Y_0 R_{i0} R_{f0} R_{h1} + Y_1 R_{i1} R_{f0} R_{h0} + Y_2 R_{i0} R_{f1} R_{h0} + Y_3 R_{i1} R_{f1} R_{h1} \qquad (17)$$

Alternatively, to get an XNOR gate, we can follow a method similar to the XOR gate. Let us suppose we have a system similar to the one in figure 3. Again, we first need to transform the superposition in equation (13) to equation (14), but this time, we need to multiply $R_{i1}$ & $R_{f0}$ by $R_{h0}R_{h1}$. In equation (14), if we substitute $R_{i1}$ by $R_{i1}R_{h0}R_{h1}$ and $R_{f0}$ by $R_{f0}R_{h0}R_{h1}$, then we get the following:

$$Y_{XNOR} = Y_0 R_{i0} R_{f0} R_{h0} R_{h1} R_{h0} + Y_1 R_{i1} R_{h0} R_{h1} R_{f0} R_{h0} R_{h1} R_{h0} + Y_2 R_{i0} R_{f1} R_{h0} + Y_3 R_{i1} R_{h0} R_{h1} R_{f1} R_{h0} \qquad (18)$$

Substituting equation (10) in equation (18), gives equation (17), which is the XNOR operation. Figure 6 illustrates this operation. Alternatively, we could have multiplied $R_{i0}$ & $R_{f1}$ by $R_{h0}R_{h1}$ to get the exact same result.

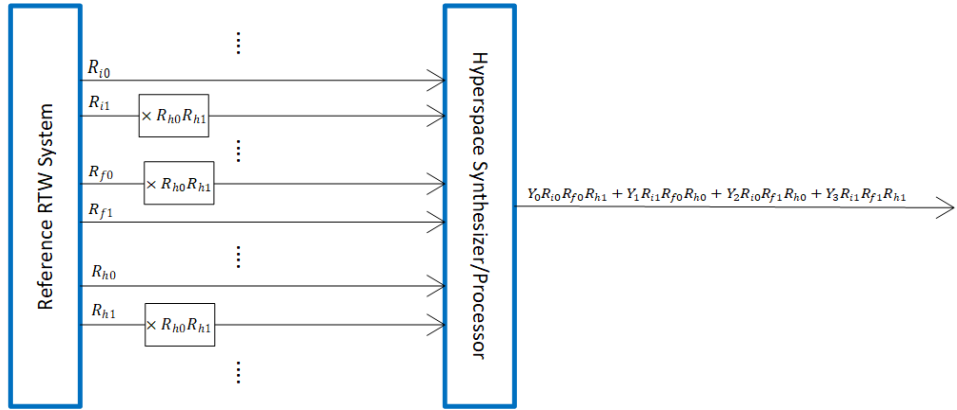

Figure 6: The full XNOR gate implementation

## 3. Conclusion

We have successfully implemented the XOR and XNOR gates that act on an exponentially large superposition by utilizing polynomial complexity, using only 4 multiplications; one for multiplying the zero value of the RTW with the one value of the RTW (example: $R_{h0}R_{h1}$ for bit $\{h\}$), and the other three multiplication happen directly on the reference wires (see figure 5 and figure 6). The XOR and XNOR operations are repeatable as well, as multiplication is a commutive operation. These gates have potential applications in challenging the supremacy of quantum computing schemes.